\documentstyle[12pt,epsf]{article}
\begin{document}
\hfill \parbox{33mm}{MS-TPI-96-17 \\
                     FAU-TP3-96/16 \\
                     hep-th/9610197}
\vspace*{10mm}
\begin{center}
\LARGE
Field Theory, Critical Phenomena and Interfaces\\[8mm]
\large\sc
Gernot M\"unster\footnote{
     Talk presented to the Graduiertenkolleg Erlangen-Regensburg
     on May~17, 1995} \\[5mm]
\normalsize
Institut f\"ur Theoretische Physik I  \\
Westf\"alische Wilhelms-Universit\"at \\
Wilhelm-Klemm-Str.~9 \\
D-48149  M\"unster, Germany
\\[8mm]
\large
Notes by {\sc H.\ Grie{\ss}hammer} and {\sc D.\ Lehmann}
\\[5mm] 
\normalsize 
Institut f\"ur Theoretische Physik III \\
Friedrich-Alexander-Universit\"at Erlangen-N\"urnberg\\
Staudtstr.~7 \\
D-91058~Erlangen, Germany
\end{center}
\setcounter{footnote}{0}
\tableofcontents

\section{Field Theory}

These lecture notes want to illustrate the close connection between
statistical mechanics and field theory not only on the formal level,
i.e.~that many concepts of one area can easily be taken over to the
other one, but also on the level of actual calculations.  To this
purpose, the last section will demonstrate that a special statistical
system, the binary fluid system, can be described by field theory in
its critical behaviour.

\subsection{Canonical Formalism}

Let us briefly review the setup of the canonical formalism of quantum
field theory.  For simplicity, we restrict ourselves to the case of a
Hermitean scalar field $\phi(x)$ in Minkowski space.

There is a Hilbert space ${\cal H}$ of physical states $| \psi \,
\rangle$, containing the {\em vacuum} $| 0 \, \rangle$ as the state of
lowest energy.  We assume that every state vector can be written as a
linear combination of products of field operators acting on the vacuum
$| 0 \, \rangle$.  Furthermore, ${\cal H}$ carries a {\em unitary
representation} $U(a,\Lambda)$ of the {\em Poincar\'e group}, where $a$
denotes a space-time translation and $\Lambda = (\Lambda^\mu\,_\nu)$ a
(homogeneous) Lorentz transformation.  The vacuum is singled out by the
fact that it is the only state vector which is {\em invariant} under
$U(a,\Lambda)$.  The generators $P_\mu$ of space-time translations,
\begin{equation}
U(a,{\rm 1\mskip-4mu l}) = \exp \left({\rm i} P_\mu a^\mu \right)\,,
\end{equation}
have a spectrum which is
confined to the closed forward light cone\footnote{We use the Minkowski 
metric $g_{\mu\nu} = {\rm diag}\,(1,-1,-1,-1)$ }
({\em ``spectral condition''})
\begin{equation}
\label{spectralcondition}
p^2 = p_{\mu}\,p^\mu \ge 0\,; \qquad p^0 \ge 0\,.
\end{equation}
$P^0$ here denotes the Hamiltonian $H$.  The scalar field $\phi(x)$ is a
Hermitean, operator-valued distribution on ${\cal H}$, which transforms
covariantly under Poincar\'e transformations:
\begin{equation}
\label{Poincaretrafo}
U(a,\Lambda)\,\phi(x)\,U^\dagger(a,\Lambda)
= \phi(\Lambda x + a)\,.
\end{equation}
The theory is quantised by imposing {\em canonical commutation
relations} between the fundamental field $\phi(x)$ and its canonical
conjugate momentum $\pi(x)$ at equal times:
\begin{equation}
\begin{array}{rcl}
\label{ETCR}
{[}\,\phi(t,{\bf x}),\pi(t,{\bf y})\,{]}  &=&
   {\rm i}\,\delta^{(3)}({\bf x}-{\bf y})\,,\\
{[}\,\phi(t,{\bf x}),\phi(t,{\bf y})\,{]} &=&
{[}\,\pi(t,{\bf x}),\pi(t,{\bf y})\,{]} = 0\,.
\end{array}
\end{equation}
Causality is guaranteed by requiring {\em locality} for the field,
i.e.~for space-like separations of the arguments the field operators
must commute with each other:
\begin{equation}
\label{locality}
[\,\phi(x),\phi(y)\,] = 0 \qquad \mbox{for} \qquad (x-y)^2 \le 0\,.
\end{equation}

As the appearance of the $\delta$-distribution in (\ref{ETCR})
indicates, local field operators as $\phi(x)$ must be viewed as
operator-valued distributions rather than functions and should be
smeared out with appropriate test functions,

\begin{equation}
\phi(f) = \int\!\!{\rm d}^4x \,\,f(x)\,\phi(x)\,.
\end{equation}
The smearing of $\phi(x)$ also in the time variable is necessary for
interacting fields.  Then, however, one can no longer postulate
well-defined commutation relations at equal times \cite{Haag}.  An
alternative to the Hilbert space formulation is to avoid operators and
states and to introduce the functional integral as an independent way
of quantising a theory.  To make contact between both approaches, we
introduce Green's functions, which play a fundamental r\^ole in both
formulations.

\subsection{Green's Functions}

Vacuum expectation values of products of field operators
\begin{equation}
\begin{array}{rcl}
{\cal W}(x_1,\dots,x_n) &:=&
\langle \, 0 | \phi(x_1)\dots\phi(x_n)| 0 \, \rangle \vspace{2ex} \\
 &=&\langle\, 0 | \phi(0,{\bf x_1}){\rm e}^{-{\rm i}(t_1-t_2)H}\dots
 {\rm e}^{-{\rm i}(t_{n-1}-t_n)H}\phi(0,{\bf x_n}) | 0 \,\rangle
\end{array}
\end{equation}
are called {\em $n$-point correlation functions} or {\em Wightman
functions}.  A more prominent r\^ole in the Minkowski space
formulation of field theory is played by {\em time-ordered Green's
functions} which are defined as the vacuum expectation values of {\em
time-ordered} products of field operators,
\begin{equation}
\begin{array}{rcl}
\label{defgreenfunction}
G(x_1,\dots,x_n) &:=&
  \langle\, 0 |{\cal T} \phi(t_1,{\bf x_1})\dots
     \phi(t_n,{\bf x_n})| 0 \,\rangle\vspace{2ex} \\
  &=&\langle\, 0 |{\cal T}\biggl\{ \phi(0,{\bf x_1})
     {\rm e}^{-{\rm i}(t_1-t_2)H}\dots {\rm e}^{-{\rm i}(t_{n-1}-t_n)H}
     \phi(0,{\bf x_n})\biggr\}| 0 \,\rangle\,;
\end{array}
\end{equation}
here ${\cal T}$ denotes the Dyson time ordering symbol,
\begin{equation}
\label{dysonT}
{\cal T} \{ \phi(x_1)\dots\phi(x_n) \}:=
  \phi(x_{\pi(1)})\dots\phi(x_{\pi(n)})
  \qquad {\rm with}\qquad
  t_{\pi(1)} \ge t_{\pi(2)}\ge \dots \ge t_{\pi(n)}\,.
\end{equation}
They can be directly related to S-matrix elements via the well-known
{\em Lehmann-Symanzik-Zimmermann (LSZ) reduction formulae}.  E.g., the
S-matrix element between an out-state with momenta $q_1, \dots, q_l$ and
an in-state with momenta $p_1,\dots ,p_n$ is given by
\begin{equation}
\begin{array}{l}
\langle\, p_1,\dots ,p_n, {\rm in} | {\rm S}
   | q_1, \dots, q_l, {\rm out} \,\rangle =  \\
\qquad = ({\rm i} Z^{-1/2})^{n+l}
   {\displaystyle \int}\!{\rm d}^4 y_1 \dots {\rm d}^4 x_l \,\,
\exp\left( {\rm i} \sum\limits_{k=1}^{n} p_k \cdot y_k
          - {\rm i}\sum\limits_{r=1}^{l} q_r \cdot x_r \right) \\
\qquad \times
\left(\Box_{y_1} + m^2\right)\cdots\left(\Box_{x_l} + m^2\right)\,\,
G(y_1,\dots ,x_l)\,,
\end{array}
\end{equation}
where $Z$ is an appropriate normalisation factor.  For more details on
the LSZ formalism the reader is referred to a standard textbook on
quantum field theory, e.g.\ Refs.~\cite{ItzyksonZuber}, \cite{Ryder}.

The most important fact about Green's functions is that they contain
all the information about the theory.  So, given either all the
Wightman or all the time-ordered correlation functions we may construct
the Hilbert space and the unitary representation $U(a, \Lambda)$ out of
them (this is the essence of the so-called {\em reconstruction
theorem}, \cite{StreaterWightman}, \cite{JaffeGlimm}). Thus, we may use
the Green's functions to {\em define} the theory, thereby avoiding the
operator formulation altogether.

\subsection{Euclidean Time Formulation}

Another important ingredient in our formulation of Euclidean field
theory is the {\em analytic continuation to imaginary times}:
\begin{equation}
x^0 \to -{\rm i} x^4, \enspace x^4 \in{\rm I\!R}\,.
\end{equation}
This is convenient for several reasons.  Generally speaking, it improves
the analytic behaviour of the various relevant functions.  In
perturbation theory it simplifies the calculation of Feynman diagrams,
e.g.\ because of the positivity of the energy denominators.  Moreover,
since in Euclidean space all distinct world-points are space-like
separated, the {\em Euclidean Green's functions} are automatically
symmetric (for bosons) in all their space-time arguments, and we do not
need any time-ordering.  Thus, in Euclidean space we have to deal with
only one kind of Green's functions and both, the Wightman and the
time-ordered Minkowski space Green's functions, may be obtained from the
Euclidean ones by appropriate analytic continuation.  Last not least,
the functional-integral formulation to be introduced in section
(\ref{functionalintegralsection}) is much better defined in the
Euclidean formulation due to the positivity of the Euclidean Green's
functions and may be interpreted as a stochastic process.

Consider the analytic continuation of the Wightman functions to complex
time arguments
$x^0_i \to x^0_i - {\rm i} x^4_i\,,\quad x_i^0, x_i^4 \in {\rm I\!R}\,$:
\begin{equation}
\begin{array}{l}
\label{continuedWightman}
{\cal W}(x^0_1 - {\rm i} x^4_1,{\bf x_1};
   \dots ;x^0_n - {\rm i} x^4_n,{\bf x_n}) = \\[3ex]
\quad = \langle\, 0 | \phi(0,{\bf x_1})
{\rm e}^{-{\rm i} (x^0_1-x^0_2)H} {\rm e}^{-(x^4_1-x^4_2)H}
\phi(0,{\bf x_2})\dots \phi(0,{\bf x_n}) | 0 \,\rangle\,.
\end{array}
\end{equation}
Due to the positivity of the Hamiltonian this is well-defined and
analytic at least for
\begin{equation}
\label{restricteddomain}
x^4_1 > x^4_2 > \dots > x^4_n \,,
\end{equation}
since we have exponential suppression in this domain. We define the
{\em Euclidean Green's functions}, also called {\em Schwinger
functions}, as the $x^0_i \to 0$ limit of the analytically continued
Wightman functions:
\begin{equation}
\label{Schwingerevolution} 
\begin{array}{rcl}
{\cal S}({\bf x_1},x^4_1; \dots; {\bf x_n}, x^4_n) &:=&
{\cal W}(-{\rm i} x^4_1,{\bf x_1}; \dots;-{\rm i} x^4_n,{\bf x_n})
= \\[3ex]
 &=&\langle\, 0 | \phi(0,{\bf x_1}){\rm e}^{-(x^4_1-x^4_2)H}\dots
      {\rm e}^{-(x^4_{n-1}-x^4_n)H}\phi(0,{\bf x_n}) | 0 \,\rangle\,.
\end{array} 
\end{equation}
For the time being this is only legitimate if the condition
(\ref{restricteddomain}) is fulfilled. It can, however, be shown
\cite{StreaterWightman} that the region of analyticity is much larger
and the Schwinger functions are actually well-defined for all
non-coinciding Euclidean world-points:
\begin{equation}
x_i \ne x_j \qquad \mbox{for} \qquad i \ne j \,.
\end{equation}
\begin{figure}[t]
\centering
\epsfysize 10cm
\centerline{\epsffile{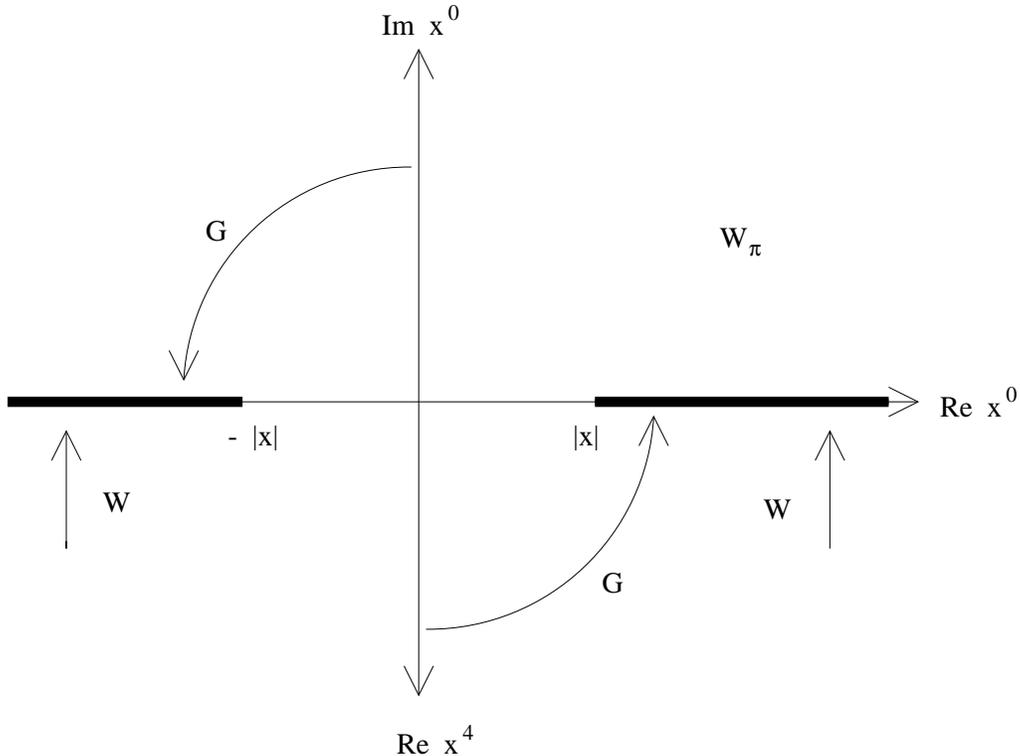}}
\parbox[t]{0.9\textwidth}{
\caption{The analyticity region of the two-point function
\protect$W(x)$ in the complex \protect$x^0$-plane.  The Wightman
function \protect$W$ is obtained by approaching the real line from
below.  The Schwinger function is defined on the imaginary
\protect$x^0$-axis.  The time-ordered Green's function \protect$G$ is
obtained from the Schwinger function by means of a Wick rotation,
indicated by curved arrows.
\label{wickrotation}}%
}
\end{figure}
Let us demonstrate this for the two-point function: The translational
invariance of the vacuum implies that ${\cal W}$ depends only on the
difference of its arguments
\begin{equation}
{\cal W}(x_1,x_2) = W(x_1-x_2) \,,
\end{equation}
which can be seen explicitly from (\ref{continuedWightman}). As a
function of complex $x^0$, $W$ is analytic in the lower half plane. For
permuted arguments we have
\begin{equation}
{\cal W}(x_2, x_1) = W_\pi(x_1-x_2) = W(x_2-x_1) \qquad \mbox{with}
\qquad W_\pi(x) := W(-x)\,,
\end{equation}
where $W_\pi$ is now analytic in the upper half plane according to
(\ref{restricteddomain}).
For space-like separations, $(x_1 - x_2)^2<0$, however, the Wightman
function is symmetric due to locality:
\begin{equation}
{\cal W}(x_1, x_2) = {\cal W}(x_2, x_1)\,,
\end{equation}
implying
\begin{equation}
W_\pi(x) = W(x)
\end{equation}
in the non-discrete region
\begin{equation}
\quad - | {\bf x} | < x^0 < | {\bf x} |
\end{equation}
of the real axis.  The well-known edge-of-the-wedge theorem then
guarantees that $W$ and $W_\pi$ form a single, analytic function in the
union of their domains, i.e.~everywhere in the complex $x^0$ plane
except along the cuts $| x^0 | > | {\bf x} |$.

The Schwinger functions and their properties have been studied in an
axiomatic setting by Osterwalder and Schrader
\cite{OsterwalderSchrader}. For our purposes the most interesting
properties are
\begin{enumerate}

\item {\bf Euclidean covariance:}
\begin{equation}
{\cal S}(\Lambda x_1+a,\dots,\Lambda x_n+a) =
{\cal S}(x_1,\dots,x_n)\,, \qquad  \Lambda \in \mbox{SO}(4)\,,
\end{equation}
where the Lorentz group of (\ref{Poincaretrafo}) is replaced by the
group of Euclidean rotations.

\item {\bf Symmetry:}
\begin{equation}
{\cal S}(x_{\sigma(1)},\dots,x_{\sigma(n)}) = {\cal S}(x_1,\dots,x_n)\,,
\end{equation}
where $\sigma$ is a permutation. This rests on the fact that distinct
Euclidean points are always space-like relative to each other.

\item {\bf Reflection positivity:}
This property replaces the Hilbert space positivity and the spectral
condition (\ref{spectralcondition}) of the Minkowskian formulation and
is necessary to guarantee that the Euclidean correlation functions may
be continued back to Minkowski space.  Formally, it is defined as
\begin{equation}
\begin{array}{l}
\displaystyle
\sum\limits_{i,j}\int\!{\rm d}^4x_1\dots{\rm d}^4x_i
   \,{\rm d}^4y_1\dots{\rm d}^4y_j\enspace
   f_i^\ast(x_1,\dots,x_i)\,f_j(y_1,\dots,y_j) \\
\qquad
\times{\cal S}(\theta x_1,\dots,\theta x_i,y_1,\dots, y_j) \ge 0\,,
\end{array}
\end{equation}
where $\theta$ is the Euclidean time reflection,
\begin{equation}
\theta ({\bf x}, x^4) := ({\bf x}, -x^4)\,,
\end{equation}
which, roughly speaking, replaces the Hermitean conjugation in
Minkowski space.
\end{enumerate}

A Euclidean quantum field theory may be defined by the set of all its
Schwinger functions. From them the Wightman functions in Minkowski
space can be reconstructed as boundary values:
\begin{equation}
{\cal W}(x_1,\dots,x_n) =
\lim \limits_{\begin{array}{c}\scriptstyle \varepsilon_i\to 0, \\
          \scriptstyle \varepsilon_i-\varepsilon_{i+1} >0  \end{array}}
{\cal S}({\bf x_1},\varepsilon_1+{\rm i} x^0_1;\dots;{\bf x_n},
   \varepsilon_n+{\rm i} x^0_n)\,,
\end{equation}
i.e.\ approaching the real $x^0$-axis from below in the case of the
two-point function, see Fig.~\ref{wickrotation}.  The advantage of the
Euclidean formulation, however, is that the Schwinger functions obey
simpler properties and are easier to handle than Wightman functions or
field operators.  In particular their symmetry is the crucial property
which opens the way to a representation in terms of functional
integrals.

The time-ordered Green's functions $G(x_1,\dots,x_n)$ may also be
obtained from the Schwinger functions, namely by approaching the real
$x^0$-axis through a counter-clockwise rotation of $\pi/2$ as
indicated in Fig.~\ref{wickrotation},
\begin{equation}
G(x_1,\dots,x_n) = \lim\limits_{\alpha \to \pi/2}
{\cal S}({\rm e}^{{\rm i} \alpha}x^0_1,\dots,
   {\rm e}^{{\rm i} \alpha}x^0_n)\,.
\end{equation}
This rule, which obviously is simpler than the one for the Wightman
functions, is called {\em Wick rotation}.

\subsection{Example: Free Field}

To illustrate the analytic continuation discussed above, let us consider
the simplest case of a {\em free} massive Hermitean scalar field
$\phi(x)$, decomposed in momentum space as \cite{Ryder}
\begin{equation}
\phi(x) = \int\!\frac{{\rm d}^3 k}{(2\pi)^3 2\omega_k}\,\,
\left[ a(k) {\rm e}^{-{\rm i} k x}
   + a^\dagger(k) {\rm e}^{{\rm i} k x} \right]\,,
\end{equation}
with $k^0 = \omega_k = \sqrt{{\bf k}^2 + m^2 }$. The annihilation and
creation operators satisfy the commutation relations
\begin{equation}
[\, a(k),a^\dagger(k') \,] = (2\pi)^3 \, 2 \omega_k \,
 \delta^3({\bf k} - {\bf k}')\,.
\end{equation}
The two-point Wightman function then is given by
\begin{equation}
W(x) = {\cal W}(x,0) = \langle\, 0 | \phi(x)\phi(0) | 0 \,\rangle
= \Delta_+(x)
= {\displaystyle \int \!\! \frac{{\rm d}^4 k}{(2\pi)^4}} \,\,
  \delta(k^2 - m^2)\, \theta(k^0) {\rm e}^{-{\rm i} k x }\,.
\end{equation}
The time-ordered two-point Green's function, the {\em Feynman
propagator}, reads
\begin{equation}
{\rm i} \Delta_{\rm F}(x) := G(x,0)
= \theta(x^0) W(x) + \theta(-x^0) W(-x)
= {\displaystyle \int \!\! \frac{{\rm d}^4 k}{(2\pi)^4} \,\,
   \frac{{\rm e}^{-{\rm i} k x}}{k^2 -m^2 + {\rm i} \varepsilon}}\,.
\end{equation}
Performing the analytic continuation of $W(x)$ yields
\begin{equation}
{\cal S}(x,0)
= \int\!{\displaystyle \frac{{\rm d}^4 k}{(2\pi)^4}} \,\,
{\displaystyle \frac{{\rm e}^{{\rm i} k x}}{k^2 +m^2}} \quad
= \,\frac{m}{(2\pi)^2 | {\bf x} |} \ {\rm K}_1(m | {\bf x} |)\,.
\end{equation}
$G(x,0)$ is obtained from ${\cal S}(x,0)$ by means of a Wick rotation of
$x^4 = {\rm i} x^0$ in a counter-clockwise sense.  In the integral we
have to rotate at the same time $k^4= {\rm i} k^0$ in a clockwise sense.
The resulting integration path in the complex $k^0$-plane bypasses the
poles at $\pm \sqrt{{\bf k}^2 + m^2}$ in the way indicated by the $+{\rm
i} \varepsilon$ prescription.  It goes from $+\infty$ to $-\infty$, and
an additional sign change has to be introduced to bring it in the form
above.  Note that the long distance behaviour of the Schwinger function
is exponential and governed by the mass $m$:
\begin{equation}
{\cal S}(x,0) \stackrel{| {\bf x} | \to \infty}{\longrightarrow}
\frac{m^2}{2 (2\pi m | {\bf x} |)^{3/2}} \
   \exp\left(-m | {\bf x} |\right)\,.
\end{equation}

\subsection{Functional Integral Formulation}
\label{functionalintegralsection}

The symmetry property of the Schwinger functions means that the
Euclidean fields commute.  In this respect they behave like classical
fields.  This allows us to interpret the $\phi(x)$'s as classical random
variables instead of quantum mechanical operators.  In this
interpretation the Schwinger functions are the $n-$point correlation
functions or {\em moments} of an appropriate probability measure
${\rm d}\mu[\phi]$:
\begin{equation}
{\cal S}(x_1,\dots,x_n) = \langle\, \phi(x_1)\dots\phi(x_n) \,\rangle
= \int\!{\rm d}\mu[\phi]\,\,\phi(x_1)\dots\phi(x_n)\,.
\end{equation}
We may define a {\em generating functional} for the Schwinger functions
as the Laplace transform of the probability measure,
\begin{equation}
\begin{array}{l}
{\cal Z}[j] := \displaystyle \biggl\langle \,
\exp\left(\int\!{\rm d}^4x\,\phi(x) j(x) \right) \, \biggr\rangle\\[3ex]
\qquad \displaystyle = \sum\limits_{n=0}^{\infty}\,\frac{1}{n!}
\int\!{\rm d}^4x_1\dots{\rm d}^4x_n\,j(x_1) \dots j(x_n)\,
\langle\,\phi(x_1)\dots\phi(x_n) \,\rangle \,,
\end{array}
\end{equation}
{}from which the moments are obtained by functional differentiation with
respect to the {\em external sources} $j(x)$:
\begin{equation}
\langle\, \phi(x_1)\dots\phi(x_n) \,\rangle
= \frac{\delta^n {\cal Z}[j]}{\delta j(x_1) \dots \delta j(x_n) }
\Bigg|_{j=0} \,.
\end{equation}
Thus the generating functional ${\cal Z}[j]$ summarises all the
information about the theory in a highly condensed form.

Formally, one may decompose ${\rm d}\mu[\phi]$ in a product of a
``Lebesgue'' measure ${\cal D}[\phi]$ on function space and a normalised
weight factor
\begin{equation}
\label{probmeasure}
{\rm d}\mu[\phi]
= \frac{1}{{\cal Z}}\,{\cal D}[\phi]\,{\rm e}^{-S_{\rm E}[\phi]} \,,
\end{equation}
where  $S_{\rm E}[\phi]$ is the Euclidean action of the field $\phi$,
\begin{equation}
\label{continuumEuclideanaction}
S_{\rm E}[\phi]
= \int\!{\rm d}^4x\,\left\{\frac{1}{2} \, \partial_\mu \phi(x)\,
   \partial_\mu \phi(x) + {\displaystyle \frac{m^2}{2}} \phi^2(x) +
   V\left(\phi(x)\right)\right\}\,.
\end{equation}
This yields the {\em functional integral formulation} of the Euclidean
correlation function and the generating functional:
\begin{eqnarray}
{\cal Z}[j] &=& {\displaystyle \frac{1}{{\cal Z}}\,\int}\!
   {\cal D}[\phi]\,\,
   \exp\left( -S_{\rm E}[\phi]+\int\!{\rm d}^4x \,\phi(x) j(x)\right)
   \,,\vspace{2ex}\\
\langle\, \phi(x_1)\dots\phi(x_n) \,\rangle
   &=& {\displaystyle \frac{1}{{\cal Z}}}
       \,{\displaystyle \int}\!{\cal D}[\phi]\,
       \phi(x_1)\dots\phi(x_n)\,{\rm e}^{ - S_{\rm E}[\phi]}\,,
\end{eqnarray}
where ${\cal Z}$ is introduced to normalise the generating functional to
${\cal Z}[0] = 1$,
\begin{equation}
{\cal Z}
= {\displaystyle \int}\!{\cal D}[\phi]\ {\rm e}^{-S_{\rm E}[\phi]}\,.
\end{equation}
However, as it stands, (\ref{probmeasure}) is a purely formal definition
since a translational invariant Lebesgue measure like ${\cal D}[\phi]$
does not exist on the infinite-dimensional function space of the
continuum field theory.

\subsection{Lattice Regularisation}

We can give a precise meaning to the functional integral expressions of
the preceeding section, if we refer to a lattice regularisation of the
Euclidean field theory.  This means, we replace the Euclidean space-time
continuum by a finite hyper-cubical lattice with lattice spacing $a$,
\begin{equation}
x_\mu = a n_\mu\,, \qquad n_\mu \enspace \mbox{integer-valued}\,.
\end{equation}
One may think of the lattice field $\phi(x)$ as the average of the
continuum field over the volume of the elementary cells of the lattice.
In momentum space, the lattice regularisation reflects itself in the
restriction of the momenta to the first Brillouin zone,
\begin{equation}
{\cal B} = \biggl\{ p_\mu \Bigg|
   -{\displaystyle \frac{\pi}{a}} \le p_\mu \le
    {\displaystyle \frac{\pi}{a}} \biggr\}\,.
\end{equation}
On a finite lattice the function space is finite-dimensional, and the
measure ${\cal D}[\phi]$ can be defined as the multi-dimensional
Lebesgue measure
\begin{equation}
{\cal D}[\phi] := \prod\limits_{x} {\rm d}\phi(x)\,,
\end{equation}
where $x$ runs over all lattice points. Space-time integrals are
replaced by sums and the differential operators $\partial_\mu$ are
replaced on the lattice by nearest-neighbour forward differences,
\begin{equation}
\begin{array}{l}
\displaystyle\int\!{\rm d}^4x\, \to \sum\limits_{x} \, a^4 \,, \\
\partial_\mu\,\phi(x) \to \Delta_\mu\phi(x) :=
    {\displaystyle \frac{1}{a}}\,\biggl(\phi(x+a \hat\mu)
    - \phi(x)\biggr)\,,
\end{array}
\end{equation}
where $\hat \mu$ is the unit vector in $\mu$-direction. This yields the
lattice analogue of (\ref{continuumEuclideanaction}):
\begin{equation}
\begin{array}{rcl}
\label{latticeaction}
S_{\rm E}[\phi] &=& \displaystyle \frac{1}{2}\sum\limits_{x}a^4
\ \biggl\{\left(\Delta_\mu\phi(x)\right)\left(\Delta_\mu\phi(x)\right)
  + {\displaystyle \frac{m_0^2}{2}}\phi^2(x) + V(\phi(x)) \biggr\} \\
 &=& \displaystyle \sum\limits_{x} a^2
  \biggl\{-\sum\limits_{\mu=1}^{4}\biggl( \phi(x)\phi(x+a\hat\mu)\biggr)
  + \biggl(4+{\displaystyle \frac{m_0^2 a^2}{4}}\biggr)\phi^2(x)
  + {\displaystyle \frac{a^2}{2}} V(\phi(x))
  \biggr\} \,.
\end{array}
\end{equation}
The lattice functional integral version of the Schwinger functions,
\begin{equation}
\langle\, \phi(x_1)\dots\phi(x_n) \,\rangle
= \frac{1}{{\cal Z}}\, \int\prod_x {\rm d}\phi(x) \
\phi(x_1)\dots\phi(x_n) \ {\rm e}^{-S_{\rm E}[\phi(x)]}\,,
\end{equation}
may also be derived from the expression (\ref{Schwingerevolution}) by
discretising the Hamiltonian, breaking up the exponential into a product
of evolution operators between two neighbouring time-slices and
inserting complete sets of field eigenstates at every time-slice -- in
the same fashion as the path integral is derived in ordinary quantum
mechanics, see e.g.\ Refs.~\cite{Muenster}, \cite{Ryder},
\cite{Roepstorff}.

The continuum theory is obtained back from the lattice version by taking
two independent limits:
\begin{enumerate}
\item {\bf  Thermodynamic limit}: taking the lattice size and
              -- with it -- the number of degrees of freedom to
              infinity;
\item {\bf  Continuum limit}: making, loosely speaking, the lattice
         structure vanish, i.e.\ by taking the lattice spacing $a$ to
         zero while keeping physical quantities like the mass gap at
         their actual value. We will see later that the continuum limit
         of a lattice field theory corresponds to a statistical lattice
         model approaching one of its critical points.
\end{enumerate}

\subsection{Analogy to Statistical Mechanics}

The functional integral formulation of the generating functional of
Euclidean lattice field theory,
\begin{equation}
{\cal Z}[j] = \frac{1}{{\cal Z}}\,\int\prod_x {\rm d}\phi(x) \
{\rm e}^{-S_{\rm E}[\phi(x)]+\int\!{\rm d}^4x\,\phi(x)j(x)}\,,
\end{equation}
exhibits a close analogy to the statistical partition function of a
system of spin variables (or magnetic moments, in general: a local
order parameter) $\phi(x)$ on a crystal, coupled via next-neighbour
interactions (see (\ref{latticeaction})) and additionally to an external
source $j(x)$ (e.g.\ magnetic field).  The probability weight
${\rm e}^{-S_{\rm E}[\phi]}$ corresponds to the Boltzmann factor
${\rm e}^{-\beta H}$.  The vacuum expectation value of the field
\begin{equation}
\langle\, \phi(x) \,\rangle
\end{equation}
corresponds to the mean magnetisation $M$ per site of a ferromagnet, and
the two-point function
\begin{equation}
\langle\, \phi(x_1)\phi(x_2) \,\rangle
\end{equation}
is equal to the spin-spin correlation function. The (dimensionless)
correlation length $\xi$, which describes the spatial extent of
fluctuations in a physical quantity about its average, governs the
exponential decay of the correlation function in the long distance
limit,
\begin{equation}
\langle\, \phi(x_1) \phi(x_2) \,\rangle
\sim {\rm e}^{- | x_1-x_2 | /a \xi}\,.
\end{equation}
It is related to the {\em mass gap} (inverse Compton length)
$m = E_1 - E_0$ of the field theory by
\begin{equation}
\label{corrlengthana}
\xi = \frac{1}{a m}\,,
\end{equation}
as can be seen from the expansion (\ref{Schwingerevolution})
\begin{equation}
\langle\, \phi(x) \phi(0) \,\rangle
= \sum\limits_{n >0} | \langle\, 0 |\phi(0) | n \,\rangle |^2
  \, {\rm e}^{- x^4 (E_n - E_0) } \,,
\end{equation}
where we inserted a complete set of energy eigenstates and, for
simplicity, have chosen ${\bf x}$ = 0. In the long (Euclidean) time
limit the two-point function consequently decays exponentially
\begin{equation}
\langle\, \phi(x) \phi(0) \,\rangle \sim {\rm e}^{-(E_1-E_0) x^4}\,.
\end{equation}
As is well-known in statistical mechanics, the {\em magnetic
susceptibility} is related to the correlation function by
\begin{equation}
\chi = \sum\limits_{x} \langle\, \phi(x) \phi(0) \,\rangle
\end{equation}
and therefore equals the propagator at zero momentum.

\begin{table}
\centering
\begin{tabular}{|r|l|}
\hline
\rule[-1.5ex]{0ex}{4.5ex}
QUANTUM FIELD THEORY & CLASSICAL STATISTICAL MECHANICS \\[1ex]
\hline\hline
\rule[-1.5ex]{0ex}{4.5ex}
scalar field $\phi(x)$ & spin variable, local order parameter $\phi(x)$
   \\[1ex]
generating functional ${\cal Z}$ & partition function ${\cal Z}$
   \\[1ex]
Euclidean action $S_{\rm E}$ & Hamiltonian $\beta H$ \\[1ex]
vacuum expectation value $\langle\, \phi(x) \,\rangle$ &
   mean magnetisation $M$ \\[1ex]
Schwinger function ${\cal S}(x_1,x_2)$ & spin-spin correlation function
   $\langle\, \phi(x_1) \phi(x_2) \,\rangle$ \\[1ex]
inverse mass $\displaystyle\frac{1}{m}$ & correlation length $a \xi$
   \\[1ex]
Lagrangean formulation in $d$ dimensions & Hamiltonian formulation in
   $d+1$ dimensions\\[2ex]
\hline
\end{tabular}
\caption{Analogy between Euclidean quantum field theory and statistical
   mechanics
\label{analogy}
}
\end{table}

The analogy between Euclidean quantum field theory on a lattice and
statistical mechanics - summarised in Tab.~\ref{analogy} - has turned
out to be very useful.  Many concepts and methods of statistical
mechanics have been applied to field theory, and conversely, the field
theoretic renormalisation group is an important tool in statistical
mechanics.  We will make extensive use of this analogy in the second
part of this talk.

\section{Critical Phenomena}

\subsection{Critical Points}

We now turn to the question of a continuum limit.  If a continuum limit
with a finite physical mass exists, it means that by a suitable choice
of the bare parameters we can approach a limit where $a$ goes to zero
while $m$ remains finite.  According to (\ref{corrlengthana}) the
(dimensionless) correlation length has to diverge in that case.

A point in the coupling constant space -- the space whose coordinates
are the parameters or coupling constants of a theory --, where $\xi$
diverges and where the first derivatives of the relevant thermodynamic
potential, say the Gibbs potential, exist, is called a {\rm critical
point} or a {\em second order phase transition}.  For most systems, the
behaviour of many quantities {\em near} a critical point is governed by
simple power laws, the so-called {\em scaling behaviour}.  As the
temperature $T$ approaches its critical value $T_c$, the correlation
length and susceptibility, for example, diverge according to
\begin{equation}
\xi \sim \Bigg |\frac{T - T_c}{T_c} \Bigg| ^{-\nu}\,, \qquad
\chi \sim \Bigg | \frac{T - T_c}{T_c}\Bigg |^{-\gamma}\,,
\end{equation}
with certain {\em critical exponents} $\nu$ and $\gamma$. The relation
$\sim$ is to be understood in the sense that
\begin{displaymath}
f(t) \sim t^\sigma \qquad \Leftrightarrow \qquad
\lim\limits_{t\to 0} \frac{\log f(t)}{\log t} = \sigma \,.
\end{displaymath}
The magnetisation in the low temperature phase vanishes like
\begin{equation}
M \sim \Bigg| \frac{T - T_c}{T_c} \Bigg|^{\beta'}\,.
\end{equation}
In four dimensions these laws are modified by logarithmic corrections.

\subsection{Universality}

The investigation of various systems of statistical mechanics near their
critical points has revealed the property of universality.  To be
precise, the systems fall into a relatively small number of {\em
universality classes}.  The members of a class show identical critical
behaviour in the sense that their critical exponents as well as certain
other universal quantities are equal.  As an example we mention that
water at its triple point falls in the same universality class as a
ferromagnet at the Curie point, which shows off in the same critical
exponent $\nu = 0.630$.

The universality classes are distinguished by
\begin{enumerate}
  \item the number of dimensions of space (or Euclidean space-time),
  \item the number of degrees of freedom of the microscopic field,
  \item the symmetries of the system.
\end{enumerate}
Universality means that the long-range properties of a critical system
do not depend on the details of the microscopic interaction.  In
particular, also the size of the lattice spacing becomes unimportant
for the large-distance behaviour of correlation functions, if the
correlation length is large.  According to the {\em scaling
hypothesis}, the correlation length is the only relevant length scale
for the system near criticality.  This hypothesis leads to various
relations between critical exponents, such that only two independent
exponents remain.

\subsection{Renormalisation Group}

Scaling theory and universality have found a theoretical basis in the
Kadanoff-Wilson {\em renormalisation group}.  Here we shall try to give
only a brief sketch of the basic ideas of the Kadanoff-Wilson
renormalisation group.  For a more detailed presentation see e.g.\
Ref.~\cite{Goldenfeld}.

The original action $S$ with a cut-off $\Lambda$ (e.g.\ the boundary
of the Brillouin zone, $\frac{\pi}{a}$) is considered to be embedded in
an infinite-dimensional space of actions
\begin{equation}
S = \sum\limits_{i} K_i \, S_i \,,
\end{equation}
where the coefficients $K_i$ are called coupling constants and
\begin{equation}
S_i = \sum\limits_{x} {\cal L}_i
\end{equation}
with so-called {\em local operators} ${\cal L}_i$. They are functions
depending on the fields at points $x$ and a finite number of points
near $x$, and having a certain engineering dimension, e.g.\
$\phi^n(x)$ has dimension $n$. A renormalisation group transformation
$R_\lambda$ is a mapping
\begin{equation}
R_\lambda: \, S \mapsto S^{(\lambda)}
\end{equation}
in this space such that both $S$ and $S^{(\lambda)}$ describe the same
physics at large distances, but the cut-off $\Lambda$ gets lowered by a
factor $\lambda > 1$:
\begin{equation}
\Lambda \to \frac{1}{\lambda} \,\Lambda \,.
\end{equation}
In other words, $S^{(\lambda)}$ is obtained from $S$ by integrating out
degrees of freedom with high momenta near the cut-off. $R_\lambda$ can
be described in terms of the changes of coefficients
\begin{equation}
K_i \to K_i^{(\lambda)}\,.
\end{equation}

\begin{figure}[t]
\centering
\epsfxsize 13cm
\centerline{\epsffile{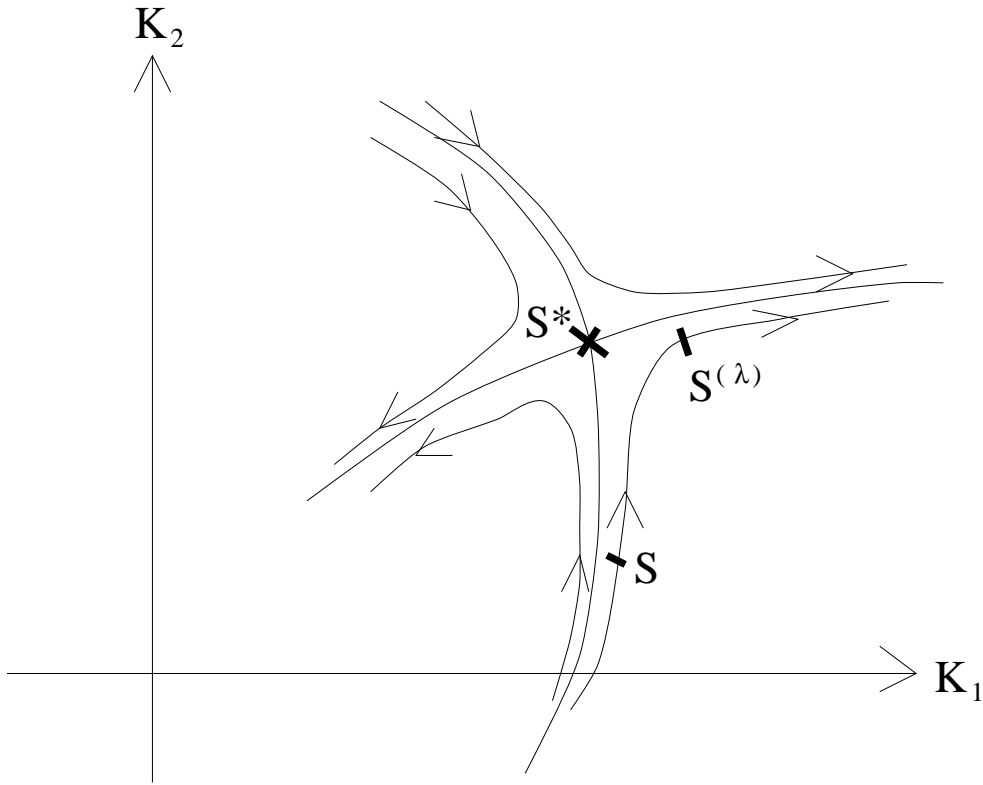}}
\parbox[t]{0.9\textwidth}{
\caption{ Vicinity of a fixed point \protect$S^\ast$ in coupling
constant space (shown are only two couplings \protect$K_1$,
\protect$K_2$ out of infinitely many) with one relevant (i.e.\ leaving
the fixed point) and one irrelevant direction (approaching the fixed
point).  The action of a renormalisation group transformation
\protect$R_\lambda$ mapping an action \protect$S$ into another action
\protect$S^{(\lambda)}$ is also shown.
\protect\label{blockspin}}%
}
\end{figure}

The most important points in the space of actions are the {\em fixed
points} $S^\ast$,
\begin{equation}
R_\lambda S^\ast = S^\ast\,,
\end{equation}
in particular those where the correlation length is infinite. Near a
fixed point, the action of $R_\lambda$ can in general be linearised and
diagonalised such that in a suitable basis
\begin{equation}
K_\alpha = K_\alpha^\ast + \delta K_\alpha
\end{equation}
it reads
\begin{equation}
K_\alpha^{(\lambda)} = K_\alpha^\ast
+ \lambda^{d_\alpha} \delta K_\alpha\,.
\end{equation}

Those terms with negative {\em scaling dimension} $d_\alpha$ get
suppressed after repeated application of the renormalisation group
transformation and are called {\em irrelevant}, since their presence
does not affect the long-distance physics.  The terms with positive
$d_\alpha$, which are a few in general, are {\em relevant}.  The values
of the corresponding coefficients $K_\alpha$ are decisive for the
long-distance physics. Terms with $d_\alpha= 0$ are called {\em
marginal}.

In this picture universality emerges in the following way.  Two
original actions $S'$ and $S''$, which belong to the {\em domain of
attraction} of the same fixed point, are mapped under the action of the
renormalisation group into the neighbourhood of the same
low-dimensional manifold
\begin{equation}
S = S^\ast + \sum\limits_{{\rm relevant}\ \alpha} K_\alpha S_\alpha \,,
\end{equation}
where for simplicity we assume that no marginal operators are present.
The critical behaviour is then determined only by the few relevant
operators in the vicinity of the fixed point.  In particular, it can be
shown that the critical indices are simple algebraic combinations of the
dimensions $d_\alpha$ belonging to them.  Thus the fixed points of the
renormalisation group determine the universality classes of the actions.

In four dimensions, perturbative calculations indicate that for the
scalar field theory under consideration there are only two relevant
operators, which are essentially the mass term $\phi^2(x)$ and the
linear term $\phi(x)$, which appears when an external ``magnetic'' field
is present.  The quartic self-interaction $\phi^4(x)$ is marginal, but
its coupling $g$ decreases under the renormalisation group
transformation.  The associated fixed point therefore has $g=0$ and
belongs to a free field theory.  It is called the {\em Gaussian fixed
point}.  Indications by non-perturbative methods lead to the result that
the Gaussian fixed point is with high certainty the only fixed point for
this theory.

\section{Interface Tensions of Binary Fluid Systems}

That universality and the renormalisation group are not just
mathematical constructs but can be tested experimentally is demonstrated
in the following example.  It exhibits again the connection between
statistical mechanics in $d+1$ dimensions and field theory, especially
the Euclidean functional integral formulation of it, in $d$ dimensions.

\subsection{Phenomenology of Binary Fluid Systems}

Consider the mixing and separation of two fluids:  Trying to mix
Cyclo-hexane ({\boldmath $C_6 H_{12}$}) and Aniline ({\boldmath $C_6 H_5
NH_2$}), one notes that below a critical temperature of $T_c=30.9^0$C,
both fluids separate spontaneously into two pure phases consisting of
Cyclo-hexane and Aniline respectively, one on top of the other.  The
surface tension $\tau$ of the interface between them vanishes as the
temperature $T$ approaches $T_c$, and one measures a scaling law for the
reduced interface tension ($k$ is Boltzmann's constant)
$\sigma:=\tau/kT$,
\begin{eqnarray}
\label{sigma}
\sigma \sim \sigma_0 t^{\mu}\,, & & \mu = 1.26\pm0.01\\
  &\displaystyle t:=\Bigg|\frac{T-T_c}{T_c}\Bigg|\;\;.&\nonumber
\end{eqnarray}
$\sigma_0$ is the critical amplitude of the interface tension.
``$\sim$'' denotes the critical behaviour of the interface tension near
$T_c$, as defined earlier.

Above $T_c$, the two fluids mix perfectly, and hence a homogeneous phase
is prepared.  Approaching the critical temperature from above, the
experimentalist notes that the mixed phase becomes less and less
transparent, and at the critical point, the system is completely opaque,
indicating that its correlation length $\xi^+$ diverges (the ``${}^+$''
denotes approach of $T_c$ from above).  No latent heat is set free, and
the system hence exhibits a second order phase transition.  The
correlation length is measured to scale above $T_c$ like
\begin{equation}
\xi^+=\xi^+_0 t^{-\nu}\;\;,\;\;\nu=0.630\pm0.002\;\;.
\end{equation}
Below $T_c$, the same critical behaviour is expected but with a
different critical amplitude $\xi_0^-$ for the correlation length.
(In this section $\xi$ is considered to be dimensionful.)

There are many other binary fluid systems like Isobutyric acid and
water, Triethylamine and water, and also systems of one fluid and one
gas, which show the same behaviour. Although the critical amplitudes
$\sigma_0\;,\; \xi^{\pm}_0$ vary considerably from system to system, the
critical exponents agree within a few percent and obey Widom's scaling
law \cite{9Int}
\begin{equation}
\label{munu}
\mu=2\nu\,.
\end{equation}
The scaling hypothesis indeed predicts $\mu$ and $\nu$ to be universal
and also gives Widom's scaling law. Renormalisation group calculations
also predict the correct value for the critical exponent $\nu$
\cite{7Int}. Binary fluid systems therefore seem to obey the scaling
hypothesis, and the critical exponents suggest that they lie in the
same universality class as the three-dimensional Ising model. As a
consequence of this, and although the critical amplitudes vary
considerably from system to system, the dimensionless quantities
\begin{equation}
\label{erpm}
R_{\pm}:=\sigma_0\left(\xi_0^{\pm}\right)^2
\end{equation}
should be universal, and experimentally this is indeed found to be the
case:
\begin{equation}
\label{exp}
R_+^{\rm exp}=0.38\pm0.02\;\;.
\end{equation}
Because $\xi_-$ is hard to measure, $R_-$ is not easily accessible
experimentally. In contradistinction, it turns out in field theory that
the low temperature value is easier to obtain than $R_+$. Monte Carlo
data of Mon \cite{18NPBII} yield
\begin{equation}
R_+^{\rm MC}=0.36\pm0.01\;\;,
\end{equation}
confirming the scaling hypothesis. In this calculation, finite-size
effects on  $\sigma$ have been shown to play an important r\^ole. On the
other hand, a first field theoretic treatment gave an unacceptable value
of $R_+=0.20$ \cite{16/17NPBII}. Is there a real discrepancy between
experiment and field theory?

\subsection{Field Theory of Binary Fluid Systems}

In the framework of field theory, one investigates critical phenomena
of the systems discussed above in the context of a Euclidean, massive
and real $\Phi^4$-theory, which is believed to be in the same
universality class as the Ising model and the binary fluid systems. One
may motivate this as follows: One needs a local order parameter which
vanishes above $T_c$ and is nonzero below $T_c$, indicating the strength
of the symmetry breaking. The difference between the concentrations of
the two fluids $A$ and $B$ at a point is surely a good candidate,
\begin{equation}
\Phi({\bf x})\propto\rho_A({\bf x})-\rho_B({\bf x})\,,
\end{equation}
and symmetry breaking is -- as in a ferromagnet -- achieved
spontaneously.

\begin{figure}[t]
\centering
\epsfxsize 13cm
\centerline{\epsffile{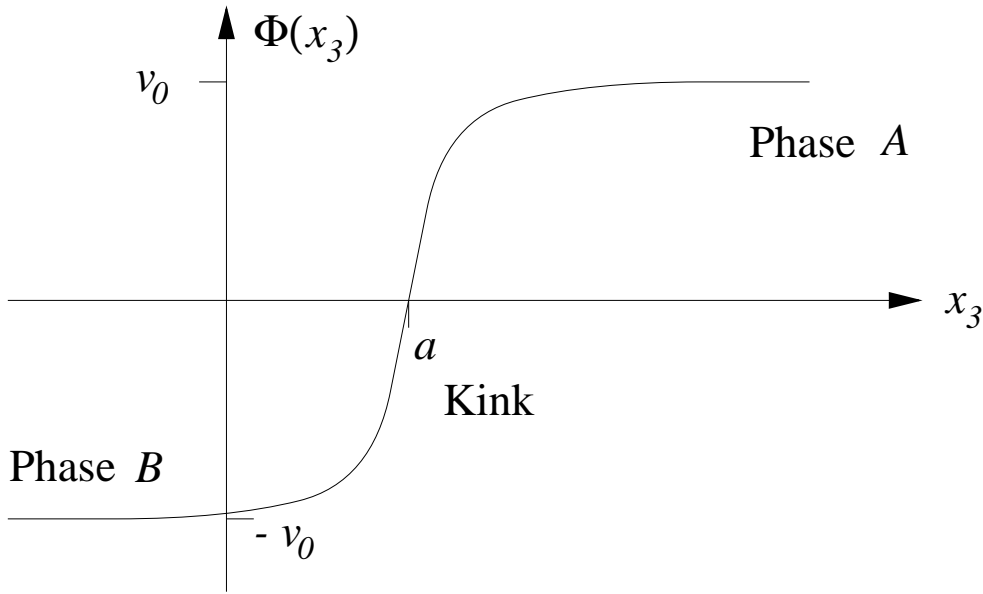}}
\parbox[t]{0.9\textwidth}{
\caption{The kink solution in \protect$\Phi^4$-theory.
\protect\label{kink}}%
}
\end{figure}

One therefore considers in field theory the bare Lagrangean
\begin{equation}
{\cal L} = \frac{1}{2}(\partial_{\mu}\Phi)(\partial^{\mu}\Phi) + V(\Phi)
\end{equation}
with the double well potential
\begin{equation}
V(\Phi)=-\frac{1}{4}m_0^2\Phi^2 +\frac{1}{4}g_0\Phi^4
+\frac{3}{8}\frac{m_0^4}{g_0}
=\frac{1}{4!} g_0 \left(\Phi^2-v_0^2\right)^2 \,,
\end{equation}
whose minima in the phase with broken symmetry ($v_0^2>0$) lie at
\begin{equation}
\Phi_0=\pm v_0 = \pm \sqrt{\frac{3m_0^2}{g_0}}\ .
\end{equation}
One of the minima, $\Phi=v_0$, is identified with the system being in
the $A$-phase, the other one, $\Phi=-v_0$, with the $B$-phase.
In the phase with spontaneously broken symmetry, $T<T_c$, when the two
components of the fluid separate, one therefore obtains the picture of
Fig.~\ref{kink} for $\Phi$, when one moves on a trajectory
perpendicular to the interface between the two fluids. The interface has
been taken to be perpendicular to the $x_3$-direction, the Euclidean
time. It corresponds to the well-known kink in field theory, upon which
will be dwelled in more detail below.

Mind that $m_0\,,\,g_0$ are bare quantities and need renormalisation.
The {\it renormalised} mass is -- as above -- the inverse of the second
moment correlation function in the low temperature phase,
\begin{equation}
\label{masscorr}
m_R=\frac{1}{\xi^-}\,,
\end{equation}
and hence vanishes at the critical point.

One may keep in mind that we are not interested in the solution of this
model, in a mass spectrum etc., but only in its behaviour near the
critical point.  It is the great advantage of universality that we are
spared a detailed comparison between $\Phi^4$-theory and a statistical
model of the binary fluid system.  In order to make contact with
experiment, one has to extract the analogue of the interface tension
$\sigma$ in $\Phi^4$-theory.  A suitable definition comes from
considering tunneling in a finite volume.  Consider a cylinder-type
geometry, where the Euclidean space is a square of side-lengths $L$ and
periodic boundary conditions, while the Euclidean time $x_3$ remains
non-compact.  The Hamiltonian of $\Phi^4$-theory in $2+1$ dimensions is
the generator of translations in the $x_3$- (i.e.~time-) direction.  In
the infinite volume limit $L\to\infty$, the symmetry $\Phi\to -\Phi$ of
the Hamiltonian is spontaneously broken at low temperatures, since the
field acquires a nonzero vacuum expectation value
\begin{equation}
\langle\, \Phi \,\rangle=\pm v_0\,,
\end{equation}
and two different but energetically degenerate ground states
\begin{equation}
\langle\, + | \Phi | + \,\rangle = +v_0 \;\; \mbox{ or }\;\;
\langle\, - | \Phi | - \,\rangle = -v_0
\end{equation}
exist. On the other hand spontaneous symmetry breaking does not occur in
a finite volume: Rather, the degeneracy between the two ground states is
lifted by tunneling effects. There exists a unique vacuum
$| S \,\rangle$, which has energy $E_{0S}=0$ and is symmetric,
\begin{equation}
\langle\, S | \Phi | S \,\rangle=0\,,
\end{equation}
and another antisymmetric state $| A \,\rangle$ with an energy
$E_{0A}$, which in the infinite volume limit approaches zero. This is
of course exactly the result of the WKB approximation of the double
well potential in ordinary quantum mechanics in the case that the wall
between the wells is very high. Sidney Coleman's presentation
\cite{Coleman} of the double well in field theory is still unbeaten,
and for details of the following presentation, the reader may consult
his lectures.

Symmetric and antisymmetric state are in the finite volume to lowest
order given by
\begin{eqnarray}
| S \,\rangle &=& \frac{1}{\sqrt{2}}
\left( | + \,\rangle + | - \,\rangle\right) \nonumber\\
| A \,\rangle &=& \frac{1}{\sqrt{2}}
\left( | + \,\rangle- | - \,\rangle\right)\,.
\end{eqnarray}
The tunneling rate is given by the transfer matrix sandwiched between
the states $\langle\, + |$ and $| \pm \,\rangle$, $T$ denoting here and
in the following Euclidean time:
\begin{equation}
\label{trans}
\langle\, + | {\rm e}^{-HT} | \pm \,\rangle
= \frac{1}{2}\left(1 \pm {\rm e}^{-E_{0A}T}\right)\,.
\end{equation}
I will now substantiate that the energy of the lowest antisymmetric
state vanishes exponentially with increasing $L$, and that it is
related to the interface tension by (see Refs.~\cite{Fisher},
\cite{PFisher}, \cite{Brezzn})
\begin{equation}
\label{pfisch}
E_{0A}= C{\rm e}^{-\sigma L^2}\,.
\end{equation}
The transition amplitude in the functional integral formulation is
\begin{equation}
\label{transamp}
\lim\limits_{T\to\infty} \langle\, + | {\rm e}^{-HT} | - \,\rangle
= \int {\cal D}\Phi\;{\rm e}^{-S[\Phi]}
\end{equation}
with the boundary conditions
\begin{equation}
\Phi({\bf x})\to\left\{\begin{array}{cc}v_0 , & \; x_3=+\infty\\
                                     -v_0 , & \; x_3=-\infty
                       \end{array}\,.\right.
\end{equation}
In the semiclassical (WKB) approximation the functional integral is
expanded around the configurations of least action. The classical kink
solution to the equations of motion,
\begin{equation}
\label{class}
\Phi_c({\bf x})=\sqrt{\frac{3m_0^2}{g_0}}\tanh\frac{m_0}{2}(x_3-a)\,,
\end{equation}
has the smallest action of all configurations interpolating between
$+v_0$ and $-v_0$. Here, $a$ is the free parameter which specifies the
location of the kink on the $x_3$-axis. The classical energy of the kink
is related to the interface tension,
\begin{equation}
\label{treel}
S_c= 2\frac{m_0^3}{g_0}L^2 =\sigma_0 L^2 \,,
\end{equation}
in this approximation since the reduced interface tension $\sigma$ of
the statistical system is exactly the free energy of the kink per unit
surface, i.e.~its action per unit surface in the Euclidean field theory.

\begin{figure}[t]
\centering
\epsfxsize 13cm
\centerline{\epsffile{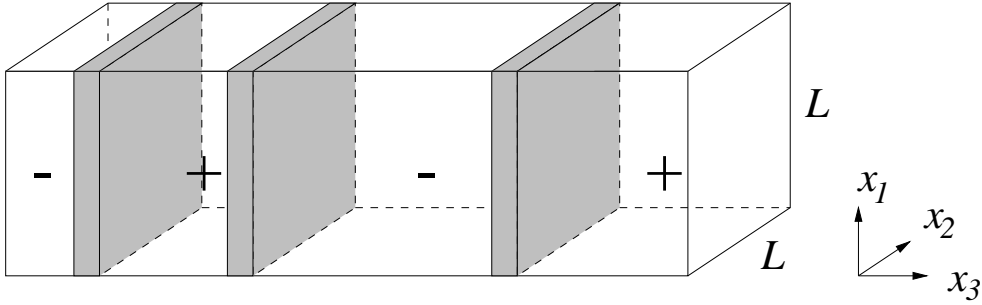}}
\parbox[t]{0.9\textwidth}{
\caption{A typical, dominating 3-kink-contribution to the transition
amplitude (\protect\ref{transamp}) of the matrix element
$\protect\langle\, + | {\rm e}^{-HT} | - \,\rangle$
at a temperature just below the critical one. Shaded regions show the
location of the kink, the \protect$\pm$-signs in between the adopted
vacuum state.
\protect\label{box}}
}
\end{figure}

As shown in Fig.~\ref{box}, one has to take into account that the
system can tunnel several times between the two states $| + \,\rangle$
and $ |- \,\rangle$, and in the WKB approximation, each of these
tunneling processes will (when only tunneling takes long enough because
of a sufficient barrier height) contribute to the functional integral a
factor
\begin{equation}
K{\rm e}^{-S_c}T\,.
\end{equation}
The factor $K$ can be calculated, but I shall not discuss it here.

Therefore, having to cross an even (odd) number of interfaces in time
development from a state
$| + \,\rangle$ ($| - \,\rangle$) to $| + \,\rangle$, one can estimate
(\ref{trans}) by
\begin{eqnarray}
\langle\, + | {\rm e}^{-HT} | \pm \,\rangle & \sim &
   \sum\limits_{n\; {\rm even(odd)}} K^n {\rm e}^{-nS_c}
              \frac{T^n}{n!} = \nonumber\\
 & & = \exp\left(K{\rm e}^{-S_c}T\right)\pm
                      \exp\left(-K{\rm e}^{-S_c}T\right)\,.
\end{eqnarray}
The factor $T^n/n!$ comes from integration over all {\em different}
locations of the kinks, i.e.~taking into account double counting.
Comparison with (\ref{trans}) indeed shows now (\ref{pfisch}),
\begin{equation}
\label{surftens}
E_{0A}= K {\rm e}^{-S_c}=C{\rm e}^{-\sigma L^2}
\end{equation}
in $3$ dimensions, so that one knows now how to obtain the interface
tension in the finite volume: It can be calculated from the energy
splitting between the ground state and the first excited state. For
$L\to\infty$, $E_{0A}\to 0$ and spontaneous symmetry breaking is
possible, all tunneling processes being exponentially suppressed.
In the following, we adopt (\ref{surftens}) as definition of the
interface tension in field theory.

\subsection{Calculating $R_+$ in Field Theory}

This section outlines the work of references \cite{Mue89} and
\cite{Mue90}.  In three-dimensional field theory, the coupling $g_0$
has a positive mass dimension and -- albeit the theory is hence
super-renormalisable -- infrared divergencies forbid one to construct
the critical, i.e.~massless, theory perturbatively.  One way out is to
consider the theory in $4-\epsilon$ dimensions and extrapolate to the
case $\epsilon =1$, but it was this technique which led to the
unacceptable value of $R_+=0.20$ in an ${\cal
O}(\epsilon^2)$-calculation \cite{16/17NPBII}.  Also it turned out that
convergence was very poor, the ${\cal O}(\epsilon)$-calculation giving
$R_+=0.14$.

Another technique \cite{7Int}, \cite{20Int} made renormalisation group
calculations in $d=3$ possible by renormalised perturbation theory:
Adopting a renormalisation scheme in which the renormalised coupling is
denoted $g_R$, the perturbative expansion goes in terms of the
renormalised, dimensionless variable
\begin{equation}
u_R:=\frac{g_R}{m_R}\,.
\end{equation}
Renormalised expansions in terms of $u_R$ do not show an infrared
problem since even at the critical point, where $m_R\to 0$, $u_R\to
u_R^*$ remains finite. Thus, information about the critical theory can
be obtained by perturbation theory at $u_R^*$, but $u_R$ is not a
parameter but fixed by the fixed point value $u_R^*$, neither is it
small.

In our case we choose the renormalised coupling as
\begin{equation}
g_R:=3\frac{m_R^2}{v_R^2}\,.
\end{equation}
The fixed point value of $u_R$ is then
\begin{equation}
u_R^*=15.1\pm 1.3
\end{equation}
{}from various analytical results in the literature.

Clearly, as one approaches the fixed point, i.e.~the point of transition
between mixed and separated phase, the interface tension vanishes and
quantum fluctuations should become important. They are the large entropy
fluctuations of statistical mechanics. The saddle point approximation of
the functional integral takes into account the quadratic fluctuations
about the classical kink solution (\ref{class})
\begin{eqnarray}
\Phi({\bf x}) & = & \Phi_c({\bf x})+\eta({\bf x})\nonumber\\
\rightarrow & & S=S_c+\frac{1}{2}\int d^3x\;
   \left[\eta({\bf x})\,M\,\eta({\bf x}) +{\cal O}(\eta^3)\right]
\end{eqnarray}
with
\begin{equation}
M = -\partial_{\mu}\partial^{\mu} + m_0^2
- \frac{3}{2}m_0^2\cosh^{-2}\frac{m_0}{2}(x_3-a)
\end{equation}
and corresponds to a one-loop perturbative calculation.

The (lengthy) calculation is for the two-dimensional case outlined in
Coleman's lectures \cite{Coleman}, and in the three-dimensional case it
has been performed in Ref.~\cite{Mue90}. The result is that the
fluctuations modify the energy of the antisymmetric state to
\begin{equation}
E_{0A} = 2\left( \frac{S_c^{1/2}}{2\pi}\right)
\left | \frac{\det'M}{\det M_0} \right|^{-1/2}\;{\rm e}^{-S_c}\,,
\end{equation}
where
\begin{equation}
     M_0= -\partial_{\mu}\partial^{\mu}+m_0^2 \,,
\end{equation}
and all contributions from multi-kink configurations have again been
taken into account.  Because $M$ has zero eigenvalues corresponding to
translations of the kink in time (parameter $a$), the determinant of $M$
is calculated without the zero modes, as indicated by the prime over the
determinant.  The zero modes have to be treated separately by the method
of collective coordinates \cite{37NPB} and give rise to the factor
$S_c^{1/2}\propto L$.

The determinant can be calculated exactly for flat interfaces using heat
kernel and $\zeta$-function techniques \cite{Mue89}.  Three types of
contributions of the final result are worth noting:  First, it produces
the counter-terms which are required to convert the unrenormalised
parameters $m_0\,,\,g_0\,,\,v_0$ of (\ref{class}) into renormalised
ones, giving secondly an additional factor $1/L$.  Finally, it gives a
substantial one-loop correction to the term proportional to $L^2$ in the
exponent, and hence to the interface tension (\ref{surftens}).

Omitting the details of the calculation of the determinant as well, one
arrives by comparison with (\ref{surftens}) at an interface tension
$\sigma$, which has a negligible, exponentially small $L$-dependence and
is in the infinite volume limit given by a Laurent series in $u_R$,
namely
\begin{equation}
\sigma = 2\frac{m^2}{u_R} \left[1-\frac{u_R}{4\pi}\left(\frac{39}{32}-
      \frac{15}{16}\ln 3\right) + {\cal O}(u_R^2) \right]\,.
\end{equation}
Comparison with the tree level result (\ref{treel}) shows that the
correction is not negligible. With (\ref{sigma}/\ref{munu}/\ref{erpm})
and (\ref{masscorr}), one finally obtains
\begin{eqnarray}
R_- & = & \sigma \left(\xi^-\right)^2 =
         \sigma_0\left(\xi^-_0\right)^2 = \nonumber\\
 & = & \frac{2}{u_R^*} \left[1-\frac{u_R^*}{4\pi}\left(\frac{39}{32}-
      \frac{15}{16}\ln 3\right) + {\cal O}(u_R^{*2}) \right]\\
 & = & 0.1024\pm 0.0088\,.\nonumber
\end{eqnarray}
The one-loop contribution amounts to $22\%$ of the leading term.

As noted at the beginning, experimentalists have measured $R_+$,
i.e.~above the critical temperature, while we calculated $R$ below
$T_c$, but with the help of another universal quantity, the conversion
factor \cite{3040NPB}
\begin{equation}
\frac{\xi_0^+}{\xi_0^-}=1.96\pm0.03\,,
\end{equation}
one finally obtains
\begin{equation}
R_+ = 0.39\pm0.03
\end{equation}
in good agreement with the experimental value (\ref{exp}). A
more recent calculation of my Diplom-student Siepmann gave
$u_R^*=14.73\pm0.14$, leading to $R_- = 0.1057\pm0.0010$.
This number is in excellent agreement with recent Monte Carlo
calculations by Hasenbusch, Pinn and coworkers \cite{Pinn}, which
resulted in
\begin{equation}
    R_- = 0.1056\pm0.0019\,.
\end{equation}


\end{document}